# Beltrami States for Plasma Dynamics Models

B.K. Shivamoggi*
J.M. Burgers Centre and Fluid Dynamics Laboratory
Department of Physics
Eindhoven University of Technology
5600 MB Eindhoven, The Netherlands

**Abstract**

The various plasma models -

- incompressible magnetohydrodynamic (MHD) model
- compressible MHD model
- incompressible Hall MHD model
- compressible Hall MHD model
- electron MHD model
- compressible Hall MHD with electron inertia model

notwithstanding the diversity of the underlying physics, are shown to exhibit some common features in the Beltrami states like -

- certain robustness with respect to the plasma compressibility effects (albeit in the barotropy assumption);
- the *Bernoulli* condition.

The Beltrami states for these models are deduced by minimizing the appropriate total energy while keeping the appropriate total helicity constant. A Hamiltonian formulation framework is used to carry out these variational problems.

---

*Permanent Address: University of Central Florida, Orlando, FL 32816-1364.

# 1 Introduction

It is well known that a significant class of exact solutions of the equations of plasma dynamics emerges under the Beltrami condition - the local current density is proportional to the magnetic field - the *force-free* state (Lundquist [1], Lust and Schluter [2]). These Beltrami solutions are also known to correlate well with real plasma behavior (Priest and Forbes [3], Schindler [4]). The purpose of this paper is to consider several models of plasma dynamics -

- incompressible magnetohydrodynamic (MHD) model
- compressible MHD model
- incompressible Hall MHD model
- compressible Hall MHD model
- electron MHD model
- compressible Hall MHD with electron inertia model,

which have quite diverse underlying physics. Nevertheless, the isotopological energy-lowering Beltramization process is shown to induce exhibition of some common features by these plasma models in the final Beltrami states. The Beltrami states are deduced using a Hamiltonian formulation framework of the various plasma models.

# 2 The Non-canonical Hamiltonian Formulation

Hamiltonian formulations have traditionally played an important role in both classical and quantion mechanics of particles and fields. However, Hamiltonian formulations were not introduced into fluid plasma problems until recently. This is due to the fact that the Eulerian variables in fluid-flow problems are non-canonical (Salmon [5]) because they are not related to the corresponding Lagrangian variables by a canonical transformation. Besides the Hamiltonian system in question is infinite-dimensional because the dynamical variables are now fields and functions of state become functionals of state.

For an illustration of the non-canonical Hamiltonian formulation framework used in the various plasma dynamics cases in this paper let us consider the incompressible hydrodynamic model described by Euler's equations (in usual notations),

$$\frac{\partial \boldsymbol{\omega}}{\partial t} = \nabla \times (\mathbf{v} \times \boldsymbol{\omega}) \tag{1a}$$

$$\nabla \cdot \mathbf{v} = 0 \tag{1b}$$

where $\boldsymbol{\omega}$ is the vorticity -

$$\boldsymbol{\omega} \equiv \nabla \times \mathbf{v}. \tag{2}$$

Equations (1a,b) have the Hamiltonian formulation -

$$\frac{1}{2} \int_V \boldsymbol{\psi} \cdot \boldsymbol{\omega} \, dV \tag{3a}$$

where V is the volume occupied by the fluid and

$$\mathbf{v} \equiv \nabla \times \boldsymbol{\psi}. \tag{3b}$$

Here, we have put $|\boldsymbol{\psi}| = 0$ on the boundary $\partial V$, and $\boldsymbol{\psi}$ is made unique by imposing the gange condition -
$$\nabla \cdot \boldsymbol{\psi} = 0. \qquad (4)$$

We choose $\omega$ to be canonical variable, and following Olver [6], take
$$J = -\nabla \times (\boldsymbol{\omega} \times (\nabla \times (\cdot))) \qquad (5)$$

as an $\boldsymbol{\omega}$-dependent differential operator which produces a skew-symmetric transformation of vector functions vanishing on $\partial V$ and satisfies a closure condition on an associated symplectic two-form (see [6] for the proof).

The Hamilton equation is then
$$\frac{\partial \boldsymbol{\omega}}{\partial t} = J \frac{\delta H}{\delta \boldsymbol{\omega}} \qquad (6)$$

which is just equation (1a). Here, $\delta H/\delta \boldsymbol{\omega}$ is the variational derivative. Thus, in the Hamiltonian formulation, the Euler equations characterize geodesic flow on an infinite-dimensional group of volume-preserving diffeomorphisms.

The operator $J$ may be seen to induce the Poisson bracket (Shepherd [7]) -
$$[F, G] = (\frac{\delta F}{\delta \boldsymbol{\omega}}, J \frac{\delta G}{\delta \boldsymbol{\omega}})$$

$$= -\int_V \frac{\delta F}{\delta \boldsymbol{\omega}} \cdot \nabla \times [\boldsymbol{\omega} \times (\nabla \times \frac{\delta G}{\delta \boldsymbol{\omega}})] dV$$

$$= -\int_V (\nabla \times \frac{\delta F}{\delta \boldsymbol{\omega}}) \cdot [\boldsymbol{\omega} \times (\nabla \times \frac{\delta G}{\delta \boldsymbol{\omega}})] dV$$

$$= \int_V \boldsymbol{\omega} \cdot [(\nabla \times \frac{\delta F}{\delta \boldsymbol{\omega}}) \times (\nabla \times \frac{\delta G}{\delta \boldsymbol{\omega}})] dV \qquad (7)$$

which is defined on admissible functionals $F[\boldsymbol{\omega}]$ and $G[\boldsymbol{\omega}]$ satisfying
$$\nabla \cdot \begin{pmatrix} \frac{\delta F}{\delta \boldsymbol{\omega}} \\ \frac{\delta G}{\delta \boldsymbol{\omega}} \end{pmatrix} = \boldsymbol{0} \text{ in } V \text{ and } \left|\frac{\delta F}{\delta \boldsymbol{\omega}}\right|, \left|\frac{\delta G}{\delta \boldsymbol{\omega}}\right| = 0 \text{ on } \partial V. \qquad (8)$$

(7) satisfies the anti-symmetry property -
$$[F, G] = -[G, F] \qquad (9a)$$

and the Jacobi identity,
$$[[F, G], K] + [[G, K], F] + [[K, F], G] = 0 \qquad (9b)$$

and other required algebraic properties of Poisson brackets.

The dynamics underlying equations (1a,b) can then be represented symplectically by
$$\frac{\partial F}{\partial t} = [F, H] \qquad (10)$$

for admissible functionals $F[\omega]$.

# 3 The Incompressible MHD Model

Let us now consider the MHD incompressible model as a preamble to discussion of several more complex models of plasma dynamics. The original Hamiltonian formulation of the MHD model was given by Morrison and Greene [8] (see also Morrison [9]). The MHD equations (in the usual notations),

$$\frac{\partial \boldsymbol{\omega}}{\partial t} = \nabla \times (\mathbf{v} \times \boldsymbol{\omega}) + \frac{1}{\rho c} \nabla \times (\mathbf{J} \times \mathbf{B}) \tag{11}$$

$$\frac{\partial \mathbf{A}}{\partial t} = \mathbf{v} \times (\nabla \times \mathbf{A}) \tag{12}$$

have the Hamiltonian formulation

$$H \equiv \frac{1}{2} \int_V \left( \boldsymbol{\psi} \cdot \boldsymbol{\omega} + \frac{1}{c} \mathbf{A} \cdot \mathbf{J} \right) dV. \tag{13}$$

Here $\rho$ is the constant mass density of the plasma, and

$$\rho \mathbf{v} \equiv \nabla \times \boldsymbol{\psi}, \quad \mathbf{B} \equiv \nabla \times \mathbf{A}, \quad \boldsymbol{\omega} \equiv \nabla \times \mathbf{v} \tag{14}$$

V being the volume occupied by the plasma.

The Hamilton equations are then

$$\begin{pmatrix} \frac{\partial \boldsymbol{\omega}}{\partial t} \\ \frac{\partial \mathbf{A}}{\partial t} \end{pmatrix} = J \begin{pmatrix} \frac{\delta H}{\delta \boldsymbol{\omega}} \\ \frac{\delta H}{\delta \mathbf{A}} \end{pmatrix} \tag{15}$$

where

$$J \equiv \begin{pmatrix} -\nabla \times (\frac{\boldsymbol{\omega}}{\rho} \times (\nabla \times (\cdot))) & -\nabla \times (\frac{\mathbf{B}}{\rho} \times (\cdot)) \\ -\frac{\mathbf{B}}{\rho} \times (\nabla \times (\cdot)) & 0 \end{pmatrix}. \tag{16}$$

The operator $J$ may again be viewed to induce a *Poisson bracket* as shown in Section 2.

The *Casimir* invariants for this problem are annihilators of the Poisson brackets which become degenerate when expressed in terms of these *natural* quantities. The Casimir invariants are therefore solutions of the equations -

$$J \begin{pmatrix} \frac{\delta \mathcal{C}}{\delta \boldsymbol{\omega}} \\ \frac{\delta \mathcal{C}}{\delta \mathbf{A}} \end{pmatrix} = \begin{pmatrix} \mathbf{0} \\ \mathbf{0} \end{pmatrix}. \tag{17}$$

It may be verified that two such solutions are

$$\begin{pmatrix} \frac{\delta \mathcal{C}_{(1)}}{\delta \boldsymbol{\omega}} \\ \frac{\delta \mathcal{C}_{(1)}}{\delta \mathbf{A}} \end{pmatrix} = \begin{pmatrix} \mathbf{0} \\ \mathbf{B} \end{pmatrix}. \tag{18}$$

or
$$\mathcal{C}_{(1)} = \int_V \mathbf{A} \cdot \mathbf{B} \, dV \tag{19}$$

and
$$\begin{pmatrix} \frac{\delta \mathcal{C}_{(2)}}{\delta \boldsymbol{\omega}} \\ \frac{\delta \mathcal{C}_{(2)}}{\delta \mathbf{A}} \end{pmatrix} = \begin{pmatrix} \mathbf{A} \\ \boldsymbol{\omega} \end{pmatrix} \tag{20}$$

or
$$\mathcal{C}_{(2)} = \int_V \boldsymbol{\omega} \cdot \mathbf{A} \, dV = \int_V \mathbf{v} \cdot \mathbf{B} \, dV. \tag{21}$$

$\mathcal{C}_{(1)}$ is the total magnetic helicity while $\mathcal{C}_{(2)}$ is the total cross helicity.

The invariance of the total magnetic helicity is related to the conservation of the degree of knottedness of magnetic field lines (Moffatt [10]). The invariance of the total cross helicity is related to the conservation of the degree of mutual knottedness of vortex lines and magnetic field lines - this remains intact even though the vortex lines are no longer frozen in the plasma in the MHD model (Moffatt [10]).

The total magnetic helicity and the total cross helicity are not positive definite, so one cannot imagine developments of the minimum magnetic helicity or the minimum cross helicity states. Actually, one obtains a Beltrami state by minimizing $H$ while keeping $\mathcal{C}_{(1)}$ fixed,
$$\frac{\delta H}{\delta \mathbf{A}} = \lambda_{(1)} \frac{\delta \mathcal{C}_{(1)}}{\delta \mathbf{A}} \tag{22}$$

or
$$\frac{1}{c}\mathbf{J} = \lambda_{(1)} \mathbf{B} \tag{23}$$

which is the *force-free* state (Woltjer [11]).

On the other hand, minimization of $H$, keeping $\mathcal{C}_{(2)}$ fixed, leads to another Beltrami state -
$$\frac{\delta H}{\delta \boldsymbol{\omega}} = \lambda_{(2)} \frac{\delta \mathcal{C}_{(2)}}{\delta \boldsymbol{\omega}} \tag{24}$$

or
$$\boldsymbol{\psi} = \lambda_{(2)} \mathbf{A} \tag{25a}$$

or
$$\rho \mathbf{v} = \lambda_{(2)} \mathbf{B} \tag{25b}$$

which is the *Alfvénic* state (Hasegawa [12]). The MHD equations (1) and (2) become linear in this state. (25b) further implies (on taking the scalar product of both sides with a material surface element $\mathbf{S}$) that constancy of ion mass flux through $\mathbf{S}$ and constancy of magnetic flux through $\mathbf{S}$ become equivalent in the Beltrami state.

Combining (23) and (25), we obtain for the Beltrami state,
$$\frac{1}{c}\mathbf{J} = \frac{\lambda_{(1)}}{\lambda_{(2)}} \rho \mathbf{v}. \tag{26}$$

Further, for the Beltrami state given by both (23) and (25), we obtain the *Bernoulli* condition -
$$\frac{p}{\rho} + \frac{1}{2}\mathbf{v}^2 = \text{const}, \quad \forall \mathbf{x} \in V \tag{27}$$
as in the incompressible hydrodynamic case.

# 4 The Compressible MHD Model

In compressible MHD the plasma density $\rho$ is no longer constant and evolves according to the mass conservation equation -
$$\frac{\partial \rho}{\partial t} + \nabla \cdot (\rho \mathbf{v}) = 0. \tag{28}$$

The plasma pressure field now plays a dynamical role and is no longer a passive variable as in incompressible MHD where it simply adjusts instantaneously so as to keep the velocity and magnetic fields solenoidal. This necessitates closing the compressible MHD equations by adding an equation of state. This may be accomplished by assuming the plasma to be barotropic, i.e., the plasma pressure is a single-valued function of the plasma density (see Morrison [13]) for a full-fledged compressible MHD development).

Assuming the barotropic condition[1]
$$\nabla P \equiv \frac{1}{\rho} \nabla p \tag{29}$$
the compressible MHD equations
$$\frac{\partial \boldsymbol{\omega}}{\partial t} = \nabla \times (\mathbf{v} \times \boldsymbol{\omega}) + \frac{1}{c} \nabla \times (\mathbf{J} \times \frac{\mathbf{B}}{\rho}) \tag{30}$$
$$\frac{\partial \mathbf{A}}{\partial t} = \mathbf{v} \times (\nabla \times \mathbf{A}) \tag{12}$$
have the Hamiltonian formulation,
$$H = \frac{1}{2} \int_V (\boldsymbol{\psi} \cdot \boldsymbol{\omega} + \frac{1}{c} \mathbf{A} \cdot \mathbf{J}) \, dV \tag{31}$$
where, we stipulate as in (14),
$$\rho \mathbf{v} \equiv \nabla \times \boldsymbol{\psi} \tag{32}$$
but $\rho$ is not a constant. However, (32) in conjunction with equation (28), implies the restrictive condition $\partial \rho / \partial t = 0$, implying that the density is frozen in time.[2]

---

[1] Barotropic behavior typically results on assuming that either the specific entropy or the temperature is constant in space and time - P then represents the specific enthalpy.

[2] This is generally valid if the plasma pressure exceeds the magnetic pressure.

The Hamilton equations are then

$$\begin{pmatrix} \frac{\partial \boldsymbol{\omega}}{\partial t} \\ \frac{\partial \mathbf{A}}{\partial t} \end{pmatrix} = J \begin{pmatrix} \frac{\delta H}{\delta \boldsymbol{\omega}} \\ \frac{\delta H}{\delta \mathbf{A}} \end{pmatrix} \tag{15}$$

where,

$$J \equiv \begin{pmatrix} -\nabla \times \left( \left( \frac{\boldsymbol{\omega}}{\rho} \right) \times (\nabla \times (\cdot)) \right) & -\nabla \times \left( \frac{\mathbf{B}}{\rho} \times (\cdot) \right) \\ -\frac{\mathbf{B}}{\rho} \times (\nabla \times (\cdot)) & \mathbf{0} \end{pmatrix}. \tag{33}$$

(33) looks formally the same as (16), but is different because it acts on a different class of functions.

The *Casimir* invariants for this problem are solutions of the equations -

$$J \begin{pmatrix} \frac{\delta \mathcal{C}}{\delta \boldsymbol{\omega}} \\ \frac{\delta \mathcal{C}}{\delta \mathbf{A}} \end{pmatrix} = \begin{pmatrix} \mathbf{0} \\ \mathbf{0} \end{pmatrix}. \tag{34}$$

It may again be verified that two such solutions are

$$\mathcal{C}_{(1)} = \int_V \mathbf{A} \cdot \mathbf{B} \, dV \tag{19}$$

$$\mathcal{C}_{(2)} = \int_V \boldsymbol{\omega} \cdot \mathbf{A} \, dV = \int_V \mathbf{v} \cdot \mathbf{B} \, dV. \tag{21}$$

It is to be observed that the conservation of total magnetic helicity does not require barotropic conditions unlike the conservation of total cross helicity, which is predicated on the existence of the barotropy condition (29).

Minimization of $H$, keeping $\mathcal{C}_{(1)}$ fixed, gives the *force-free* state

$$\frac{1}{c}\mathbf{J} = \lambda_{(1)} \mathbf{B} \tag{23}$$

while minimization of $H$, keeping $\mathcal{C}_{(2)}$ fixed, gives

$$\boldsymbol{\psi} = \lambda_{(2)} \mathbf{A} \tag{35a}$$

or

$$\rho \mathbf{v} = \lambda_{(2)} \mathbf{B} \tag{35b}$$

which is the generalized *Alfvénic* state pertinent for a compressible plasma.

Combining (23) and (35), we obtain for the Beltrami state,

$$\frac{1}{c}\mathbf{J} = \frac{\lambda_{(1)}}{\lambda_{(2)}} \rho \mathbf{v} \tag{36}$$

which is the same as that, namely (26), in the incompressible case. This is plausible because, under the restrictive condition $\partial \rho / \partial t = 0$ implicit in (32), the vector fields $\mathbf{J}$ and $\rho \mathbf{v}$ are solenoidal for both incompressible and compressible MHD Beltrami states. Thus, the MHD Beltrami states exhibit characterizations that seem to show certain robustness with respect to the plasma compressibility effects (albeit in the barotropy assumption).

Further, in the Beltrami state given by (23) and (35), we obtain the *Bernoulli* condition,

$$P + \frac{1}{2}\mathbf{v}^2 = \text{const}, \quad \forall \mathbf{x} \in V \tag{37}$$

as in the compressible hydrodynamic case (Shivamoggi and van Heijst [14]).

## 5 The Incompressible Hall MHD Model

In a high-$\beta$ plasma, on length scales in the range $d_e < \ell < d_i$, where $d_s \equiv c/\omega_{ps}$, $s = i, e$, is the skin depth, the electrons decouple from the ions and this results in an additional transport mechanism for the magnetic field via the Hall current (Sonnerup [15]), which is the ion-inertia contribution in Ohm's law (see equation (34) below). The Hall effect leads to the generation of whistler waves whose -

- frequency lies between ion-cyclotron and electron-cyclotron frequencies $\omega_{ci}$ and $\omega_{ce}$, respectively,
- phase velocity exceeds that of Alfvén waves for wavelengths parallel to the applied magnetic field less than $d_i$.

Further, the decoupling of ions and electrons in a narrow region around the magnetic neutral point (where the ions become unmagnetized while the electrons remain magnetized[3]) allows for rapid electron flows in the ion dissipation region and hence a faster magnetic reconnection process in the Hall MHD regime (Mandt et al. [18]).

The incompressible Hall MHD equations (which were actually formulated by Lighthill [19] long ago following his far-sighted recognition of the importance of the Hall term in the generalized Ohm's law),

$$\frac{\partial \mathbf{\Omega}_i}{\partial t} = \nabla \times (\mathbf{v}_i \times \mathbf{\Omega}_i) \tag{38}$$

$$\frac{\partial \mathbf{A}}{\partial t} = \frac{1}{c}\mathbf{v}_i \times \mathbf{B} - \frac{1}{nec}\mathbf{J} \times \mathbf{B} \tag{39}$$

where $n$ is the constant number density of ions (or electrons) and $\mathbf{\Omega}_i$ is the generalized vorticity,

$$\mathbf{\Omega}_i \equiv \boldsymbol{\omega}_i + \boldsymbol{\omega}_{ci}, \ \boldsymbol{\omega}_i \equiv \nabla \times \mathbf{v}_i, \ \boldsymbol{\omega}_{ci} \equiv \frac{e\mathbf{B}}{m_i c} \tag{40}$$

have the Hamiltonian formulation,

$$H \equiv \frac{1}{2}\int_V [\boldsymbol{\psi}_i \cdot \mathbf{\Omega}_i + \frac{1}{c}\mathbf{A} \cdot (\mathbf{J} - ne\mathbf{v}_i)] \ dV. \tag{41}$$

---

[3]The Hall MHD model has a superficial resemblance to the Landau [16] model of superfluid $^4He$ which considers the superfluid below the lambda point as a two-fluid system - a frictionless superfluid and a viscous normal fluid. The vorticies in superfluid $^4He$ are frozen in the superfluid but slip past the normal fluid (Donnelly [17]).

where,
$$m_i n \mathbf{v}_i \equiv \nabla \times \boldsymbol{\psi}_i. \tag{42}$$

Here, we have considered an incompressible, two-fluid, quasi-neutral plasma and have neglected the electron inertia.

The Hamilton equations are then
$$\begin{pmatrix} \frac{\partial \boldsymbol{\Omega}_i}{\partial t} \\ \frac{\partial \mathbf{A}}{\partial t} \end{pmatrix} = J \begin{pmatrix} \frac{\delta H}{\delta \boldsymbol{\Omega}_i} \\ \frac{\delta H}{\delta \mathbf{A}} \end{pmatrix} \tag{43}$$

where,
$$J \equiv \begin{pmatrix} -\nabla \times (\frac{\boldsymbol{\Omega}_i}{m_i n} \times (\nabla \times (\cdot))) & \mathbf{0} \\ \mathbf{0} & \frac{c\mathbf{B}}{ne} \times (\cdot) \end{pmatrix}. \tag{44}$$

The *Casimir* invariants for this problem are solutions of the equations,
$$J \begin{pmatrix} \frac{\delta \mathcal{C}}{\delta \boldsymbol{\Omega}_i} \\ \frac{\delta \mathcal{C}}{\delta \mathbf{A}} \end{pmatrix} = \begin{pmatrix} \mathbf{0} \\ \mathbf{0} \end{pmatrix}. \tag{45}$$

It may be verified that two such solutions are
$$\begin{pmatrix} \frac{\delta \mathcal{C}_{(1)}}{\delta \boldsymbol{\Omega}_i} \\ \frac{\delta \mathcal{C}_{(1)}}{\delta \mathbf{A}} \end{pmatrix} = \begin{pmatrix} \mathbf{0} \\ \mathbf{B} \end{pmatrix} \tag{46}$$

or
$$\mathcal{C}_{(1)} = \int_V \mathbf{A} \cdot \mathbf{B} \, dV \tag{47}$$

as with classical MHD, and
$$\begin{pmatrix} \frac{\delta \mathcal{C}_{(2)}}{\delta \boldsymbol{\Omega}_i} \\ \frac{\delta \mathcal{C}_{(2)}}{\delta \mathbf{A}} \end{pmatrix} = \begin{pmatrix} \frac{e\mathbf{A}}{m_i c} + \mathbf{v}_i \\ \left(\frac{e}{m_i c}\right)^2 \mathbf{B} \end{pmatrix} \tag{48}$$

or
$$\mathcal{C}_{(2)} = \int_V (\frac{e\mathbf{A}}{m_i c} + \mathbf{v}_i) \cdot \boldsymbol{\Omega}_i \, dV \tag{49}$$

which is the total generalized ion cross helicity.[4]

---

[4]On noting, alternatively,
$$\mathcal{C}_{(2)} = \int_V \left[ \left(\frac{e}{m_i c}\right)^2 \mathbf{A} \cdot \mathbf{B} + \mathbf{v}_i \cdot (2\boldsymbol{\Omega}_i - \boldsymbol{\omega}_i) \right] dV$$

Minimization of $H$, keeping $\mathcal{C}_{(1)}$ fixed, gives

$$\frac{\delta H}{\delta \mathbf{A}} = \lambda_{(1)} \frac{\delta \mathcal{C}_{(1)}}{\delta \mathbf{A}} \tag{50}$$

or, on redefining the Lagrange multiplier $\lambda_{(1)}$,

$$\frac{1}{c}(\mathbf{J} - ne\mathbf{v}_i) = \lambda_{(1)} \mathbf{B}. \tag{51}$$

On the other hand, minimization of $H$, keeping $\mathcal{C}_{(2)}$ fixed, gives

$$\frac{\delta H}{\delta \mathbf{\Omega}_i} = \lambda_{(2)} \frac{\delta \mathcal{C}_{(2)}}{\delta \mathbf{\Omega}_i} \tag{52}$$

or

$$\psi_i = \lambda_{(2)} \left( \frac{e\mathbf{A}}{m_i c} + \mathbf{v}_i \right) \tag{53a}$$

or

$$m_i n \mathbf{v}_i = \lambda_{(2)} \mathbf{\Omega}_i \tag{53b}$$

which is the generalized *Alfvénic* state. (54) implies that constancy of ion mass flux through a material surface element **S** and constancy of generalized magnetic flux (which is the magnetic flux plus the ion fluid vorticity flux) through **S** become equivalent in the Beltrami state (Shivamoggi [20]).

Combining (51) and (53), we obtain for the Beltrami state,

$$\frac{m_i}{e} \nabla \times \mathbf{B} - (\lambda_{(1)} \frac{m_i}{e} + \frac{e}{m_i c} \lambda_{(2)}) \mathbf{B} = \lambda_{(2)} \omega_i \tag{54}$$

in agreement with that given by Turner [21].

Further in the Beltrami state given by (51) and (53), we obtain again the *Bernoulli* condition -

$$\frac{p_i}{\rho} + \frac{1}{2} \mathbf{v}_i^2 = \text{const}, \quad \forall \mathbf{x} \in V \tag{27}$$

as in the incompressible hydrodynamic case.

# 6 The Compressible Hall MHD Model

Treatment of the compressible Hall MHD model, where the number density $n$ of ions (or electrons) is no longer constant, is again facilitated by assuming the barotropic conditions

$$\nabla P_{e,i} \equiv \frac{1}{nm_{e,i}} \nabla p_{e,i} \tag{55}$$

---

we obtain

$$\frac{\delta \mathcal{C}_{(2)}}{\delta \mathbf{A}} = \left(\frac{e}{m_i c}\right)^2 \mathbf{B}$$

as required, per (48).

the compressible Hall MHD equations

$$\frac{\partial \mathbf{\Omega}_i}{\partial t} = \nabla \times (\mathbf{v}_i \times \mathbf{\Omega}_i) \tag{56}$$

$$\frac{\partial \mathbf{A}}{\partial t} = \mathbf{v}_i \times \mathbf{B} - \frac{1}{e}\mathbf{J} \times \left(\frac{\mathbf{B}}{n}\right) \tag{39}$$

have the Hamiltonian formulation,

$$H \equiv \frac{1}{2}\int_V \left[\boldsymbol{\psi}_i \cdot \mathbf{\Omega}_i + \frac{1}{c}\mathbf{A}\cdot(\mathbf{J} - ne\mathbf{v}_i)\right] dV \tag{41}$$

where, as in (42),

$$m_i n \mathbf{v}_i \equiv \nabla \times \boldsymbol{\psi}_i. \tag{57}$$

but n is not a constant. However, (58), in conjunction with equation (28), again implies the restrictive condition $\partial n/\partial t = 0$.

The Hamilton equations are then

$$\begin{pmatrix} \frac{\partial \mathbf{\Omega}_i}{\partial t} \\ \frac{\partial \mathbf{A}}{\partial t} \end{pmatrix} = J \begin{pmatrix} \frac{\delta H}{\delta \mathbf{\Omega}_i} \\ \frac{\delta H}{\delta \mathbf{A}} \end{pmatrix} \tag{43}$$

where,

$$J \equiv \begin{pmatrix} -\nabla \times \left(\frac{\mathbf{\Omega}_i}{m_i n} \times (\nabla \times (\cdot))\right) & \mathbf{0} \\ \mathbf{0} & \frac{c\mathbf{B}}{ne} \times (\cdot) \end{pmatrix}. \tag{58}$$

The *Casimir* invariants for this problem are solutions of the equations,

$$J \begin{pmatrix} \frac{\delta \mathcal{C}}{\delta \mathbf{\Omega}_i} \\ \frac{\delta \mathcal{C}}{\delta \mathbf{A}} \end{pmatrix} = \begin{pmatrix} \mathbf{0} \\ \mathbf{0} \end{pmatrix}. \tag{59}$$

It may again be verified that two such solutions are

$$\mathcal{C}_{(1)} = \int_V \mathbf{A} \cdot \mathbf{B}\, dV \tag{47}$$

$$\mathcal{C}_{(2)} = \int_V \left(\frac{e\mathbf{A}}{m_i c} + \mathbf{v}_i\right) \cdot \mathbf{\Omega}_i\, dV. \tag{49}$$

Minimization of $H$, keeping $\mathcal{C}_{(1)}$ fixed, gives

$$\frac{1}{c}(\mathbf{J} - ne\mathbf{v}_i) = \lambda_{(1)}\mathbf{B} \tag{60}$$

while minimization of $H$, keeping $\mathcal{C}_{(2)}$ fixed, gives

$$\boldsymbol{\psi}_i = \lambda_{(2)}\left(\frac{e\mathbf{A}}{m_i c} + \mathbf{v}_i\right) \tag{61a}$$

or

$$m_i n \mathbf{v}_i = \lambda_{(2)} \mathbf{\Omega}_i \tag{61b}$$

which is the generalized *Alfvenic* state. (60) and (61) are in agreement with those given by Mahajan et al. [22].

Combining (60) and (61), we obtain for the Beltrami state,

$$\frac{m_i}{e} \nabla \times \mathbf{B} - (\lambda_{(1)} \frac{m_i}{e} + \frac{e}{m_i c} \lambda_{(2)}) \mathbf{B} = \lambda_{(2)} \boldsymbol{\omega}_i \tag{62}$$

which is the same as that in the incompressible case, namely (54), (Turner [21]). This is plausible because (62) is re-expressible in the form

$$\mathbf{J} = a \boldsymbol{\omega}_i + b \mathbf{\Omega}_i \tag{63}$$

the vector fields $\mathbf{J}$, $\boldsymbol{\omega}_{ci}$, and $\boldsymbol{\omega}_i$ being solenoidal for both incompressible and compressible Hall MHD Beltrami states.

Thus, the Hall MHD Beltrami states exhibit characterizations that also seem to show certain robustness with respect to the plasma compressibility effects (albeit in the barotropy assumption) in spite of the fact that the Hall MHD Beltrami states are neither *force-free* ($\mathbf{J} \neq a\mathbf{B}$) nor *Alfvénic* ($\mathbf{v}_i \neq b\mathbf{B}$)!

Further, in the Beltrami state given by (60) and (61), we obtain again the *Bernoulli* condition,

$$P_i + \frac{1}{2} \mathbf{v}^2 = \text{ const}, \forall \mathbf{x} \in V \tag{37}$$

as in the compressible hydrodynamic case [14].

## 7 The Electron MHD Model

In the MHD model, the dynamics is dominated by ions with electrons serving to shield out rapidly any charge imbalances. In electron MHD (EMHD), with $\ell \ll \rho_{si}$, $\rho_s$ being the gyroradius, on the other hand, the dynamics is dominated by electrons with the demagnetized ions serving to provide the neutralizing static background (Kingsep et al. [23], Gordeev et al. [24]). The assumptions underlying the EMHD model are $\ell \ll d_i$ and that the frequencies involved are greater than $\omega_{ci}$ and $\omega_{pi}$.

The magnetic field transport equation is

$$\frac{\partial \mathbf{B}_e}{\partial t} = \nabla \times (\mathbf{v}_e \times \mathbf{B}_e) \tag{64}$$

where $\mathbf{B}_e$ is the generalized magnetic field,

$$\mathbf{B}_e \equiv \mathbf{B} - d_e^2 \nabla^2 \mathbf{B}. \tag{65}$$

On using the electron mass conservation condition

$$\frac{\partial n_e}{\partial t} + \nabla \cdot (n_e \mathbf{v}_e) = 0 \tag{66}$$

and the barotropic condition (55), equation (64) may be rewritten as

$$\frac{D}{Dt}\left(\frac{\mathbf{B}_e}{n_e}\right) = \left(\frac{\mathbf{B}_e}{n_e} \cdot \nabla\right)\mathbf{v}_e. \tag{67}$$

On introducing the vector potential,

$$\mathbf{B} \equiv \nabla \times \mathbf{A} \tag{14}$$

with the gauge condition,

$$\nabla \cdot \mathbf{A} = 0 \tag{68}$$

imposed to render $\mathbf{A}$ unique, equation (64) leads to

$$\frac{\partial \mathbf{A}_e}{\partial t} = \mathbf{v}_e \times (\nabla \times \mathbf{A}_e) - \nabla \phi \tag{69}$$

where $\mathbf{A}_e$ is the generalized magnetic vector potential,

$$\mathbf{A}_e \equiv \mathbf{A} - d_e^2 \nabla^2 \mathbf{A} \tag{70}$$

and $\phi$ is an arbitrary scalar function. Equation (69) may be rewritten as

$$\frac{DA_{ei}}{Dt} = v_{ej}\frac{\partial A_{ej}}{\partial x_i} - \frac{\partial \phi}{\partial x_i}. \tag{71}$$

Note, from (65), (14), and (70), we have

$$\mathbf{B}_e = \nabla \times \mathbf{A}_e. \tag{72}$$

We obtain from equations (67) and (71),

$$\begin{aligned}\frac{D}{Dt}\left(\frac{\mathbf{A}_e \cdot \mathbf{B}_e}{n_e}\right) &= A_{ei}\frac{D}{Dt}\left(\frac{B_{ei}}{n_e}\right) + \frac{B_{ei}}{n_e}\frac{DA_{ei}}{Dt}\\ &= A_{ei}\left(\frac{B_{ej}}{n_e}\frac{\partial v_{ei}}{\partial x_j}\right) + \frac{B_{ei}}{n_e}\left(v_{ej}\frac{\partial A_{ej}}{\partial x_i} - \frac{\partial \phi}{\partial x_i}\right)\\ &= \left(\frac{\mathbf{B}_e}{n_e}\right) \cdot \nabla\left(\mathbf{v}_e \cdot \mathbf{A}_e - \phi\right).\end{aligned} \tag{73}$$

Suppose $S$ be a magnetic surface enclosing a volume $V$ and moving with the electron fluid; consider the total generalized electron magnetic helicity,

$$H_{eM} \equiv \int_V \mathbf{A}_e \cdot \mathbf{B}_e \, dV. \tag{74}$$

Then, on noting the mass-conservation condition for an electron fluid element,

$$\frac{D}{Dt}(n_e \, dV) = 0 \tag{75}$$

and using equation (63), we have,

$$\begin{aligned}
\frac{dH_{e_M}}{dt} &= \int_V \frac{D}{Dt}\left(\frac{\mathbf{A}_e \cdot \mathbf{B}_e}{n_e}\right) n_e \, dV \\
&= \int_V (\mathbf{B}_e \cdot \nabla)(\mathbf{v}_e \cdot \mathbf{A}_e - \phi) \, dV \\
&= \oint_S (\hat{\mathbf{n}} \cdot \mathbf{B}_e)(\mathbf{v}_e \cdot \mathbf{A}_e - \phi) \, dV = 0
\end{aligned} \qquad (76)$$

because $\hat{\mathbf{n}} \cdot \mathbf{B}_e = 0$ on $S$ ($\hat{\mathbf{n}}$ being the outward normal to $S$). Thus, we have in EMHD,

$$H_{e_M} = \text{ const.} \qquad (77)$$

## 7.1 A Lagrange Invariant

If we impose the gauge condition,

$$\phi - \mathbf{v}_e \cdot \mathbf{A}_e = 0 \qquad (78)$$

we have from equation (73),

$$\frac{D}{Dt}\left(\frac{\mathbf{A}_e \cdot \mathbf{B}_e}{n_e}\right) = 0. \qquad (79)$$

So, the generalized electron magnetic potential helicity is a Lagrange invariant,

$$\frac{\mathbf{A}_e \cdot \mathbf{B}_e}{n_e} = \text{ const.} \qquad (80)$$

The generalized electron magnetic potential helicity Lagrange invariant (80) signifies invariance of the topology of the generalized magnetic field in EMHD. The generalized electron magnetic potential helicity is therefore important for the study of topological properties of magnetic field lines in EMHD in analogy to the incompressible MHD case (Moffatt [10]) where (80) reduces to the magnetic helicity Lagrange invariant,

$$\mathbf{A} \cdot \mathbf{B} = \text{ const.} \qquad (81)$$

(81) was deduced by Kuz'min [25] via an impulse formulation of the incompressible MHD equations.

Using (78), equation (71) becomes

$$\frac{DA_{ei}}{Dt} = A_{ej} \frac{\partial v_{ej}}{\partial x_i}. \qquad (82)$$

On the other hand, if $\ell$ a vector field associated with an infinitesimal line element of the electron fluid, $\ell$ evolves according to (Batchelor [26]),

$$\left[\frac{\partial}{\partial t} + (\mathbf{v}_e \cdot \nabla)\right]\ell = (\ell \cdot \nabla)\mathbf{v}_e \qquad (83)$$

which is identical to equation of evolution of $\frac{\mathbf{B}_e}{n_e}$, namely, (67). Therefore, the field lines of $\frac{\mathbf{B}_e}{n_e}$ evolve as electron fluid line elements.

Next, if $\mathbf{S}$ is a vector field associated with an oriented material surface element of the electron fluid, $\mathbf{S}$ evolves according to (Batchelor [26])

$$\left[\frac{\partial}{\partial t} + (\mathbf{v}_e \cdot \nabla)\right](n_e \mathbf{S}) = -(\nabla \mathbf{v}_e)^T (n_e \mathbf{S}) \tag{84}$$

which is identical to the equation of evolution of $\mathbf{A}_e$, namely (82). Therefore, the field lines of $\mathbf{A}_e$ evolve as oriented electron fluid surface mass elements - the direction of $\mathbf{A}_e$ is orthogonal to the surface mass element.

These results imply that the generalized electron magnetic potential helicity invariant is simply physically equivalent to the mass conservation of the electron fluid element. One may even view this equivalence to provide a kind of inevitability to the generalized electron magnetic potential helicity invariant.

## 7.2 Variational Formulation

Consider now states resulting by minimizing the total energy,

$$E \equiv \frac{1}{2}\int_V (\mathbf{B}^2 + m_e n_e \mathbf{v}_e^2) dV = \frac{1}{2}\int_V \left(\mathbf{B}^2 + d_e^2 (\nabla \times \mathbf{B})^2\right) dV \tag{85}$$

subject to constancy of the total generalized magnetic helicity,

$$H_{eM} = \int_V \mathbf{A}_e \cdot \mathbf{B}_e \, dV = \text{const.} \tag{77}$$

In the second statement of equation (85), we have neglected the displacement current $\partial \mathbf{E}/\partial t$ and changes in $n_e$ - this is valid if $\omega \ll \omega_{pe}^2/\omega_{ce}$, and assumed the electron temperature $T_e$ is constant.

This leads to

$$\int_V [\{\mathbf{B} + d_e^2 \nabla \times (\nabla \times \mathbf{B})\} \cdot \delta \mathbf{B} + 2\mu \mathbf{B}_e \cdot \delta \mathbf{A}_e] \, dV = 0 \tag{86}$$

which may be rewritten as

$$\int_V [\nabla \times \mathbf{B}_e + 2\mu\{\mathbf{B}_e + d_e^2 \nabla \times (\nabla \times \mathbf{B}_e)\}] \cdot \delta \mathbf{A} \, dV = 0. \tag{87}$$

(87) leads to

$$d_e^2 \nabla \times (\nabla \times \mathbf{B}_e) + \frac{1}{2\mu}(\nabla \times \mathbf{B}_e) + \mathbf{B}_e = \mathbf{0} \tag{88}$$

which is a double Beltrami state in $\mathbf{B}_e$.

On the other hand, the equation of motion of the electrons, on assuming the barotropic condition (55), is

$$m_e \frac{\partial \mathbf{v}_e}{\partial t} + \frac{e}{m_e}\mathbf{E} - \mathbf{v}_e \times \mathbf{\Omega}_e = -\nabla\left(\frac{\mathbf{v}_e^2}{2} + \frac{p_e}{n_e}\right) \tag{89}$$

where,

$$\mathbf{\Omega}_e \equiv \boldsymbol{\omega}_e + \boldsymbol{\omega}_{ce}, \quad \boldsymbol{\omega}_e \equiv \nabla \times \mathbf{v}_e, \quad \boldsymbol{\omega}_{ce} \equiv -\frac{e\mathbf{B}}{m_e c}. \tag{90}$$

Upon taking the curl of equation (89), we obtain

$$\frac{\partial \mathbf{\Omega}_e}{\partial t} = \nabla \times (\mathbf{v}_e \times \mathbf{\Omega}_e) \qquad (91)$$

where we have noted,

$$\mathbf{E} = -\frac{1}{c}\frac{\partial \mathbf{A}}{\partial t}. \qquad (92)$$

The Beltrami state is then given by

$$\mathbf{\Omega}_e = a\mathbf{v}_e \qquad (93)$$

$a$ being on arbitrary function of space and time. (93) may be rewritten as

$$d_e^2 \nabla \times (\nabla \times \mathbf{B}) - a(\nabla \times \mathbf{B}) + \mathbf{B} = \mathbf{0} \qquad (94)$$

which is a double Beltrami state in $\mathbf{B}$ but is isomorphic to equation (88) given by the variational development.

Thus, the Beltrami state in EMHD continues to possess the usual variational characterization - the minimizer of energy on iso-helicity surfaces.

Further detail on this Beltrami state, like the determination of the arbitrary function $a$ in equation (93), becomes available on posing a Hamiltonian formulation of equation (91).

## 7.3 Hamiltonian Formulation

The Hamiltonian for this system is

$$H = \frac{1}{2}\int_V (m_e n_e \mathbf{v}_e^2 + \mathbf{B}^2)\, dV \qquad (95)$$

which may be rewritten as

$$H = \frac{1}{2}\int_V (\boldsymbol{\psi}_e \cdot \boldsymbol{\omega}_e + \frac{1}{c}\mathbf{A}\cdot\mathbf{J})\, dV \qquad (96)$$

where

$$m_e n_e \mathbf{v}_e \equiv \nabla \times \boldsymbol{\psi}_e. \qquad (97)$$

(97) implies the restrictive conditon $\partial n_e/\partial t = 0$ - this, as mentioned previously, is now, however, the assumption underlying the EMHD model. (96) may be rewritten as

$$H = \frac{1}{2}\int_V \left[\boldsymbol{\psi}_e \cdot \boldsymbol{\omega}_e - \frac{e}{c}\mathbf{A}\cdot(n_e\mathbf{v}_e)\right] dV \qquad (98)$$

which, on using (97), becomes

$$H = \frac{1}{2}\int_V \boldsymbol{\psi}_e \cdot \mathbf{\Omega}_e\, dV. \qquad (99)$$

We assume either that $\hat{\mathbf{n}}\cdot\mathbf{\Omega}_e = 0$ on a surface $S$ which bounds the volume $V$ and moves with the electron fluid or that $V$ is unbounded and $\mathbf{\Omega}_e$ falls away sufficiently rapidly.

We take $\mathbf{\Omega}_e$ to be the canonical variable and the skew-symmetric operator $J$ to be

$$J \equiv -\nabla \times \left[\left(\frac{\mathbf{\Omega}_e}{m_e n_e}\right) \times (\nabla \times (\cdot))\right]. \tag{100}$$

The Hamilton equation is then

$$\begin{aligned}\frac{\partial \mathbf{\Omega}_e}{\partial t} = J\frac{\delta H}{\delta \mathbf{\Omega}_e} &= -\nabla \times \left[\left(\frac{\mathbf{\Omega}_e}{m_e n_e}\right) \times (\nabla \times \boldsymbol{\psi}e)\right] \\ &= -\nabla \times \left[\left(\frac{\mathbf{\Omega}_e}{m_e n_e}\right) \times (n_e m_e \mathbf{v}_e)\right] \\ &= \nabla \times (\mathbf{v}_e \times \mathbf{\Omega}_e)\end{aligned}$$

as required (equation (91)).

The *Casimir* invariants for this system are solutions of the equation,

$$J\frac{\delta \mathcal{C}}{\delta \mathbf{\Omega}_e} = -\nabla \times \left[\left(\frac{\mathbf{\Omega}_e}{m_e n_e}\right) \times \left(\nabla \times \frac{\delta \mathcal{C}}{\delta \mathbf{\Omega}_e}\right)\right] = \mathbf{0} \tag{101}$$

from which,

$$\frac{\delta \mathcal{C}}{\delta \mathbf{\Omega}_e} = \mathbf{v}_e - \frac{e\mathbf{A}}{m_e c} \tag{102}$$

so,

$$\mathcal{C} = \int_V \left(\mathbf{v}_e - \frac{e\mathbf{A}}{m_e c}\right) \cdot \mathbf{\Omega}_e \, dV \tag{103}$$

which is simply the total generalized electron magnetic helicity given by (74).

The Beltrami state is the minimizer of $H$, keeping $\mathcal{C}$ constant, and is given by

$$\frac{\delta H}{\delta \mathbf{\Omega}_e} = \lambda \frac{\delta \mathcal{C}}{\delta \mathbf{\Omega}_e} \tag{104}$$

or

$$\boldsymbol{\psi}_e = \lambda \left(\mathbf{v}_e - \frac{e\mathbf{A}}{m_e c}\right) \tag{105}$$

or

$$m_e n_e \mathbf{v}_e = \lambda \mathbf{\Omega}_e \tag{106}$$

which is just the Beltrami state (93), with $a = \frac{m_e n_e}{\lambda}$.

Further, in the Beltrami state given by (106), we obtain again the *Bernoulli* condition,

$$P_e + \frac{1}{2}\mathbf{v}_e^2 = \text{const}, \ \forall \mathbf{x} \in V \tag{107}$$

as in the compressible electron hydrodynamic case.

# 8 The Compressible Hall MHD with Electron Inertia Model

Let us now consider compressible Hall MHD with electron inertia with length scales in the range $l < d_e$.

The equation of motion of the ions, on assuming the barotropic condition (55), is

$$\frac{\partial \mathbf{v}_i}{\partial t} - \frac{e}{m_i}\mathbf{E} - \mathbf{v}_i \times \boldsymbol{\Omega}_i = -\nabla \left(\frac{\mathbf{v}_i^2}{2} + P_i\right) \tag{108}$$

where,

$$\boldsymbol{\Omega}_i \equiv \boldsymbol{\omega}_i + \boldsymbol{\omega}_{ci}, \ \boldsymbol{\omega}_i \equiv \nabla \times \mathbf{v}_i, \ \boldsymbol{\omega}_{ci} \equiv \frac{e\mathbf{B}}{m_i c}. \tag{109}$$

Equation (108) may be rewritten as

$$\frac{\partial}{\partial t}\left(\mathbf{v}_i + \frac{e\mathbf{A}}{m_i c}\right) - \mathbf{v}_i \times \boldsymbol{\Omega}_i = -\nabla\left(\frac{\mathbf{v}_i^2}{2} + P_i\right). \tag{110}$$

Upon taking the curl of equation (110), we obtain

$$\frac{\partial \boldsymbol{\Omega}_i}{\partial t} = \nabla \times (\mathbf{v}_i \times \boldsymbol{\Omega}_i). \tag{111}$$

On the other hand, from the equation of motion of the electrons, namely, equation (89), we obtain equation (91).

From equations (91) and (111), the Beltrami state is then given by

$$\boldsymbol{\Omega}_e = a\mathbf{v}_e, \ \boldsymbol{\Omega}_i = b\mathbf{v}_i \tag{112}$$

$a$ and $b$ being arbitrary functions of space and time.

Assuming the quasi-neutrality condition,

$$n_e \approx n_i = n \tag{113}$$

we obtain from equation (112),

$$\frac{c}{e}\nabla \times \mathbf{B} - \frac{ne}{c}\left(\frac{1}{m_i b} + \frac{1}{m_e a}\right)\mathbf{B} = \frac{n}{b}\nabla \times \mathbf{v}_i - \frac{n}{a}\nabla \times \mathbf{v}_e \tag{114}$$

which reduces to equation (62), on dropping the electron contribution.

Further detail on this Beltrami state, like the determination of the arbitrary functions $a$ and $b$ in equation (112), becomes available on posing a Hamiltonian formulation of equations (91) and (111).

## 8.1 Non-canonical Hamiltonian Formulation

The Hamiltonian for this system is

$$H = \frac{1}{2}\int_V \left(m_i n \mathbf{v}_i^2 + m_e n \mathbf{v}_e^2 + \mathbf{B}^2\right) dV \tag{115}$$

which may be rewritten as

$$H = \frac{1}{2} \int_V \left( \boldsymbol{\psi}_i \cdot \boldsymbol{\omega}_i + \boldsymbol{\psi}_e \cdot \boldsymbol{\omega}_e + \frac{1}{c} \mathbf{A} \cdot \mathbf{J} \right) dV \tag{116}$$

where,

$$m_{e,i} \, n\mathbf{v}_{e,i} \equiv \nabla \times \boldsymbol{\psi}_{e,i}. \tag{117}$$

(117) again implies the restrictive condition $\frac{\partial n}{\partial t} = 0$. (116) may be rewritten as

$$H = \frac{1}{2} \int_V \left[ \boldsymbol{\psi}_i \cdot \boldsymbol{\omega}_i + \boldsymbol{\psi}_e \cdot \boldsymbol{\omega}_e + \frac{e}{c} \mathbf{A} \cdot n \left( \mathbf{v}_i - \mathbf{v}_e \right) \right] dV \tag{118}$$

which, on using (117), becomes

$$H = \frac{1}{2} \int_V \left( \boldsymbol{\psi}_i \cdot \boldsymbol{\Omega}_i + \boldsymbol{\psi}_e \cdot \boldsymbol{\Omega}_e \right) dV. \tag{119}$$

We assume that $V$ is unbounded and $\boldsymbol{\Omega}_{e,i}$ fall away sufficiently rapidly.

The Hamilton equations are then

$$\begin{pmatrix} \frac{\partial \boldsymbol{\Omega}_i}{\partial t} \\ \frac{\partial \boldsymbol{\Omega}_e}{\partial t} \end{pmatrix} = J \begin{pmatrix} \frac{\delta H}{\delta \boldsymbol{\Omega}_i} \\ \frac{\delta H}{\delta \boldsymbol{\Omega}_e} \end{pmatrix} \tag{120}$$

where,

$$J \equiv \begin{pmatrix} -\nabla \times \left[ \left( \frac{\boldsymbol{\Omega}_i}{m_i n} \right) \times (\nabla \times (\cdot)) \right] & \mathbf{0} \\ \mathbf{0} & -\nabla \times \left[ \left( \frac{\boldsymbol{\Omega}_e}{m_e n} \right) \times (\nabla \times (\cdot)) \right] \end{pmatrix}. \tag{121}$$

The *Casimir* invariants for this system are solutions of the equations,

$$J \begin{pmatrix} \frac{\delta \mathcal{C}}{\delta \boldsymbol{\Omega}_i} \\ \frac{\delta \mathcal{C}}{\delta \boldsymbol{\Omega}_e} \end{pmatrix} = \begin{pmatrix} \mathbf{0} \\ \mathbf{0} \end{pmatrix}.$$

It may be verified that two such solutions are

$$\mathcal{C}_{(1)} = \int_V \left( \mathbf{v}_i + \frac{e\mathbf{A}}{m_i c} \right) \cdot \boldsymbol{\Omega}_i \; dV \tag{122}$$

$$\mathcal{C}_{(2)} = \int_V \left( \mathbf{v}_e - \frac{e\mathbf{A}}{m_e c} \right) \cdot \boldsymbol{\Omega}_e \; dV \tag{123}$$

which are the total generalized ion and electron magnetic helicities, respectively.

The Beltrami state is the minimizer of $H$ keeping $\mathcal{C}_{(1)}$ and $\mathcal{C}_{(2)}$ constant, and is given by

$$\frac{\delta H}{\delta \boldsymbol{\Omega}_i} = \lambda_{(1)} \frac{\delta \mathcal{C}_{(1)}}{\delta \boldsymbol{\Omega}_i} \tag{124}$$

and
$$\frac{\delta H}{\delta \mathbf{\Omega}_e} = \lambda_{(2)} \frac{\delta \mathcal{C}_{(2)}}{\delta \mathbf{\Omega}_e} \tag{125}$$

which lead to
$$\boldsymbol{\psi}_i = \lambda_{(1)} \left( \mathbf{v}_i + \frac{e\mathbf{A}}{m_i c} \right) \tag{126a}$$

and
$$\boldsymbol{\psi}_e = \lambda_{(2)} \left( \mathbf{v}_e - \frac{e\mathbf{A}}{m_e c} \right) \tag{127a}$$

or
$$m_i n \mathbf{v}_i = \lambda_{(1)} \mathbf{\Omega}_i \tag{126b}$$

and
$$m_e n \mathbf{v}_e = \lambda_{(2)} \mathbf{\Omega}_e \tag{127b}$$

which is just the Beltrami state (112), with $a = \frac{nm_e}{\lambda_{(2)}}$ and $b = \frac{nm_i}{\lambda_{(1)}}$. Observe that (126b) is the same as that in the incompressible case, namely, (53).

Further, in the Beltrami state given by (126) and (127), we obtain the Bernoulli conditions,

$$P_i + \frac{1}{2}\mathbf{v}_i^2 = const, \ \forall \mathbf{x} \ \epsilon \ V \tag{128}$$

$$P_e + \frac{1}{2}\mathbf{v}_e^2 = const, \ \forall \mathbf{x} \ \epsilon \ V \tag{129}$$

as in the compressible ion/electron hydrodynamic case.

## 9 Discussion

The emergence of a significant class of *exact* solutions of equations governing several models of plasma dynamics and their correlation to real plasma behavior raises the question: Do plasmas have an intrinsic tendency towards Beltramization? Though a definitive answer of this question is not available yet, it may be of some help to note that the Beltramization process provides the means via which the plasma system in question can accomplish -

- ergodicity of the streamlines of the respective flow (Moffatt [27]);
- selective dissipation of total energy (Woltjer [11]);

Furthermore, in this paper we have seen that the Beltramization process also induces exhibition of some common features like -

- certain robustness with respect to the plasma compressibility effects (albeit in the barotropy assumption);
- the *Bernoulli* condition;

by the various plasma models in the final Beltrami states, despite quite diverse underlying physics. Similar results have been shown to be valid also in hdyrodynamics (Shivamoggi et al. [14] and [28]). Beltrami states appear to have a certain resemblance to critical phenomena aspects in condensed matter physics - at a critial point, as is well known, many of the precise details of the interactions between constituent subunits play essentially no role in determining the bulk properties of the system. The latter are determined primarily by the dimension of the system and the general symmetry properties of the constituent subunits. Consequently, totally dissimilar systems exhibit certain universal features in their behavior near the critical point (Stanley [29]).

## 10  Acknowledgments

This work was carried out when the author held a visiting research appointment at the Eindhoven University of Technology under the auspices of J.M. Burgers Centre. The author is very thankful to Professor Gert Jan van Heijst for his enormous hospitality. The author expresses his thanks to the referee for his helpful criticism. The author is also thankful to Professors Swadesh Mahajan and Leon Kamp for helpful discussions.

## References

[1] S. Lundquist: *Arkiv. Fysik* **2**, 361, (1950).

[2] R. Lust and A. Schluter: *Z. Astrophys.* **34**, 263, (1954).

[3] E.R. Priest, and T. Forbes: *Magnetic Reconnection*, Cambridge University Press, (2000).

[4] K. Schindler: *Physics of Space Plasma Activity*, Cambridge University Press, (2007).

[5] R. Salmon: *Ann. Rev. Fluid Mech.* **20**, 225, (1988).

[6] P.J. Olver: *J. Math. Anal. Appl.* **89**, 233, (1982).

[7] T.G. Shepherd: in *Topological Aspects of the Dynamics of Fluids and Plasmas*, Ed. H.K. Moffatt and A. Tsinober, Kluwer, p. 275, (1992).

[8] P.J. Morrison and J.M. Greene: *Phys. Rev. Lett.* **45**, 790, (1980).

[9] P.J. Morrison: *Phys. Plasmas* **12**, 058102, (2005).

[10] H.K. Moffatt: *J. Fluid Mech.* **35**, 117, (1969).

[11] L. Woltjer: *Proc. Nat. Acad. Sci.* **44**, 489, (1958).

[12] A. Hasegawa: *Adv. Phys.* **34**, 1, (1985).

[13] P.J. Morrison: in *Mathematical Methods in Hydrodynamics and Integrability in Related Dynamical Systems*, Ed. M. Tabor and Y. Treve, American Institute of Physics Conf. Proc. # **88**, (1982).

[14] B.K. Shivamoggi and G.J.F. van Heijst: *Phys. Lett. A* **372**, 5688, (2008).

[15] B.U.O. Sonnerup: in *Solar System Plasma Physics*, Ed. L.J. Lanzerotti, C.F. Kennel, and E.N. Parker, North Holland, p. 45, (1979).

[16] L.D. Landau: *J. Phys.* (USSR) **5**, 71, (1941).

[17] R.J. Donnelly: *Quantized Vortices in Helium II*, Cambridge Univ. Press, (1991).

[18] M.E. Mandt, R.E. Denton and J.F. Drake: *Geophys. Res. Lett.* **21**, 73, (1994).

[19] M.J. Lighthill: *Phil. Trans. Roy. Soc. (London)* **A 252**, 397, (1960).

[20] B.K. Shivamoggi: *Phys. Lett. A* **373**, 708, (2009).

[21] L. Turner: *IEEE Trans. Plasma Sci.* **PS-14**, 849, (1986).

[22] S.M. Mahajan, R. Miklaszewski, K.J. Nikol'skaya and N.L. Shatashvili: *Phys. Plasmas* **8**, 1340, (2001).

[23] A.S. Kingsep, K.V. Chukbar, and V.V. Yan'kov: *Rev. Plasma Phys.* **16**, 243, (1990).

[24] A.V. Gordeev, A.S. Kingsep, and L.I. Rudakov: *Phys. Reports* **243**, 215, (1994).

[25] G.A. Kuz'min: *Phys. Lett. A* **96**, 83, (1983).

[26] G.K. Batchelor: in *An Introduction to Fluid Dynamics*, Cambridge Univ. Press, p. 132, (1967).

[27] H.K. Moffatt: in *Whither Turbulence*, Ed. J.L. Lumley, p. 250, Springer-Verlag, New York, (1990).

[28] B.K. Shivamoggi, G.J.F. van Heijst and J.J. Rasmussen: *Phys. Lett. A* **374**, 2309, (2010).

[29] H.E. Stanley: *Rev. Mod. Phys.* **71**, S538, (1999).